# Routing with Privacy for Drone Package Delivery Systems

Geoffrey Ding, Alex Berke, Kwassi H. Degue
Hamsa Balakrishnan
Massachusetts Institute of Technology
Cambridge, MA, USA
{gding, aberke, kdegue, hamsa}@mit.edu

Karthik Gopalakrishnan
Stanford University
Stanford, CA, USA
gkarthik@stanford.edu

Max Z. Li
University of Michigan
Ann Arbor, MI, USA
maxzli@umich.edu

*Abstract*—Unmanned aerial vehicles (UAVs), or drones, are increasingly being used to deliver goods from vendors to customers. To safely conduct these operations at scale, drones are required to broadcast position information as codified in remote identification (remote ID) regulations. However, location broadcast of package delivery drones introduces a privacy risk for customers using these delivery services: Third-party observers may leverage broadcast drone trajectories to link customers with their purchases, potentially resulting in a wide range of privacy risks. We propose a probabilistic definition of privacy risk based on the likelihood of associating a customer to a vendor given a package delivery route. Next, we quantify these risks, enabling drone operators to assess privacy risks when planning delivery routes. We then evaluate the impacts of various factors (e.g., drone capacity) on privacy and consider the trade-offs between privacy and delivery wait times. Finally, we propose heuristics for generating routes with privacy guarantees to avoid exhaustive enumeration of all possible routes and evaluate their performance on several realistic delivery scenarios.

I. INTRODUCTION

Unmanned aerial vehicles (UAVs), or drones, are emerging as a way to deliver important and time-sensitive items as well as regularly purchased products such as groceries and household goods. For example, drones have been trialed as a way to deliver critical medical supplies (e.g., vaccines, personal protective equipment, and pharmaceuticals) under time constraints in less-developed regions [1, 2]. A number of large companies have also entered into preliminary agreements with external drone operators to leverage drones for package delivery. For example, Walmart has partnered with drone operator Zipline to make on-demand deliveries within an 80 kilometer radius of a Walmart store [3], Walgreens has begun trials of drone package delivery services with drone operator Wing [4], and the restaurant chain Chipotle has been testing the use of drones for food delivery [5].

There is also an emerging business model in which companies that operate fleets of drones enable package deliveries between smaller local businesses and customers. Such companies are already operating in Europe, Australia, and the United States [6, 7]. In general, our work assumes a model where a drone operator services multiple businesses.

*A. Context and motivation*

The need for safe drone operations—collision avoidance in particular—has led to regulations that require drones to publicly broadcast their positions (e.g., Remote ID regulations in the US [8, 9]). However, sharing the location of package delivery drones introduces a privacy risk: Third-party observers can collect broadcast location information and deduce sensitive parameters such as the origins (e.g., vendors) and destinations (e.g., customer addresses) of drone delivery routes. Then, third-party observers may associate customers with their purchases, leading to possible invasions of customer privacy ranging from targeted advertisement to inference of personal health information.

**Example 1** (Customer interest in privacy). A customer may order specialized products (e.g., for particular diets, lifestyles, medical conditions, or demographic characteristics) from a vendor that offers door-to-door delivery via drone. By collecting the broadcast positions of drones engaged in package delivery operations, third-party observers can ascertain which customers ordered these products. This results in a loss of customer privacy and may even endanger the personal safety of the customer.

Drone deliveries also present a risk for vendors regardless of whether or not they offer specialized products.

**Example 2** (Vendor interest in privacy). A vendor may utilize drone package delivery services to fulfill customer orders. A rival vendor that identifies the vendor's customers via broadcast drone trajectories could gain a competitive advantage by targeting advertisements and promotions to those customers. In this case, not only is the vendor at risk of losing customers, but its customers may be repelled by the targeted advertising.

The NASA University Leadership Initiative (grants #80NSSC21M0071 and #80NSSC20M0163) provided funds to assist the authors with their research, but this article solely reflects the opinions and conclusions of its authors and not any NASA entity. G. Ding was additionally supported by a National Science Foundation Graduate Research Fellowship.

In Examples 1 and 2, the overarching concern is identical: Vendors and customers both wish to prevent third-party observers from being able to deduce who ordered from whom. The risks to customers and vendors can be mitigated if the information that a particular customer received a package from a specific vendor is protected. With this in mind, we next provide an overview of solution approaches we take to reason about and reduce such privacy risks posed by drone package delivery services.

*B. Solution approach*

Given the potential privacy risks of drone package delivery services, we seek to first define such privacy risks quantitatively and then consider routing strategies that reduce such privacy risks. We motivate our privacy-preserving routing strategies and subsequent formal definitions with examples:

**Example 3** (Privacy via aggregation). Consider a setting where two different customers at addresses $a_1$ and $a_2$ order packages to be delivered from two different vendors $v_1$ and $v_2$, respectively. If a drone is routed directly from $v_1$ to $a_1$ and then from $v_2$ to $a_2$ (Fig. 1(a)), a third-party observer can easily infer which customer received a delivery from which vendor, compromising privacy. In contrast, suppose the drone first picks up the packages from the two vendors, in any order, prior to performing subsequent deliveries (Fig. 1(b)). Then, a third-party observer of the drone's trajectory would be less able to determine which item was delivered to which customer.

**Example 4** (Privacy via decoy vendors). Consider a setting where vendor $v_1$ sends a package via drone delivery to a customer at address $a_1$. If a drone is routed directly from $v_1$ to $a_1$, a third-party observer can easily infer that $a_1$ received a delivery from $v_1$. In contrast, suppose that the drone makes a "decoy stop" at vendor $v_2$ (henceforth, *decoy vendor*) prior to delivering the package from $v_1$ to $a_1$ (Fig. 1(c)). Then, an observer of the drone's trajectory would be less able to infer which vendor delivered to $a_1$. Realistically, the decoy vendor $v_2$ should be located along the path from $v_1$ to $a_1$ and could, for example, be used as a recharging location for the drone.

The improvements in privacy given by aggregation and decoy vendors do not come for free: Consider an extreme case of 100 vendor visits, of which 99 are decoy vendors, followed by one customer visit. While the chance that a third-party observer could deduce the correct vendor is only 1%, the customer would be negatively impacted by a lengthy wait time. Thus, trade-offs between privacy and efficiency should also be considered.

*C. Related work*

Previous works have explored a variety of strategies to mitigate the privacy risks of linking customers with their purchasing behaviors. For example, recent speculative work proposed various "private delivery network" models, using intermediary delivery sites and aggregating packages before final delivery, and where the different models leveraged different delivery network topologies to improve customer privacy [10]. Our work goes further by formalizing routing strategies and quantifying associated privacy loss in the context of existing real-world applications.

Other solutions to mitigate these risks include having customers swap store loyalty cards to prevent the aggregation of their individual purchasing data by large chain stores while still accessing the lower prices offered to loyalty card holders [11]. Related solutions, including our contributions in this work, fit into the framework of *obfuscation*, which includes strategies to mitigate the impact of data collection and profiling by adding noise to a data collection process, resulting in data that is more ambiguous and therefore less valuable [11]. For example, the TrackMeNot browser extension, with hundreds of thousands of downloads, was designed to achieve privacy in web search by obfuscating a user's actual searches amidst a stream of programmatically generated decoy searches [12].

While these works may seem speculative with respect to their real-world impact, related privacy projects have since been proposed by large companies such as Google. Google announced it would remove third-party cookies (which enable tracking of individual users, their purchases, and their preferences) from the popular Chrome web browser and proposed an alternative method for advertisers to reach web users that provided greater privacy [13, 14]. Google's proposed "Topics API" provides a limited number of "topics" to third parties that represent a web user's broader interests. Noise is incorporated as random, incorrect information (topics), which may be provided to the third party at any given time. In doing so, web users now have "plausible deniability" in ways similar to the obfuscation works described above. Google's announcement came after increasing pressure by privacy advocates to limit what third parties can learn about individuals. This work builds upon these ideas to address emerging privacy concerns extending from the digital realm to the physical space of customer purchases, logistics networks, and delivery modalities.

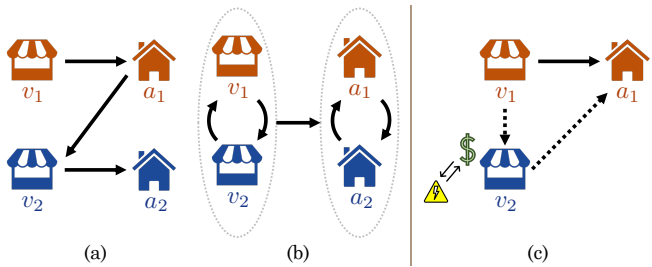

Fig. 1.  (a) Stops are made directly from vendor to customer; a third-party observer has a 100% chance of inferring which customer received a delivery from which vendor. (b) Stops are first made at the vendors in random order (randomness denoted by dashed circles), then at the customer addresses in random order; a third-party observer has a 50% chance of inferring which customer received a delivery from which vendor. (c) Two routing strategies (solid arrow versus dashed arrows) with vendors $v_1, v_2$, and customer $a_1$, where one strategy (dashed arrows) includes a "decoy stop" at vendor $v_2$.

*D. Contributions*

Our contributions in this work are as follows:
1) We propose a probabilistic definition of privacy risk and provide an algorithm to compute this privacy risk given a drone delivery route.
2) We analyze various factors (e.g., drone capacity) that impact privacy risk.
3) We provide routing heuristics to facilitate route planning with privacy considerations.
4) We examine how geography influences the trade-off between privacy and efficiency using realistic delivery scenarios.

## II. SETUP, ASSUMPTIONS, AND NOTATION

*A. Setup*

We take our setting to be a dense, urban environment where there is sufficient demand for drone delivery services. Drone operators transport orders from vendors to customers and are unrelated to either; we do not consider a vendor—such as a big-box store—that manages its own fleet of delivery drones.

The only restrictions on routes are vendor precedence (i.e., a vendor must be visited before its corresponding customer address) and order fulfillment (i.e., all customers must be visited). We explicitly *exclude* the online setting: We collect several orders, send out a drone to fulfill those orders, and amass any incoming requests in a separate route for the next available drone. Therefore, we are not concerned with dynamic rerouting of drones, nor do we consider privacy gain or loss from dynamic reordering of vendor and customer stops.

*B. Assumptions*

We assume that each customer orders from exactly one vendor and that each vendor has no more than one customer. All payloads have unit size and mass, so a drone of integral capacity $c$ can carry up to $c$ items. Since our focus is on defining, computing, and analyzing privacy risk, we consider a simplified setting and exclude considerations such as limited drone range, battery recharging, and impacts of payload on drone operation.

From the perspective of the third-party observer, we assume knowledge of (and only of) a drone's trajectory and model. The drone's trajectory is constructed from location broadcast as part of remote ID requirements. A drone must also broadcast a unique identifier, such as its serial number, that may be cross-referenced with registration information to determine a drone's model and, by extension, its capabilities (or, as relevant in this work, its capacity). The broadcast of such unique identifiers is already stipulated in existing remote ID regulations [9].

*C. Notation*

We present some notation and basic definitions that will be used throughout the rest of this work. We use $n$ as the number of package delivery orders, or number of orders for short. We denote by $v_i$ and $a_i$ the $i^{\text{th}}$ vendor and customer address, respectively, with $v_i$ delivering to $a_i$; we seek to analyze and mitigate the risk of third-party observers learning which $v_i$ delivers to an $a_j$. Sets of $n$ unique customers and $m = n + n_d$ unique vendors (with $n_d$ decoy vendors) are denoted by $A = \{a_1, \ldots, a_n\}$ and $V = \{v_1, \ldots, v_m\}$, respectively. The cardinality of a set is represented by $|\cdot|$; as such, $|A| = n$ and $|V| = m$.

A route $r$ is an ordered set of *stops*, where a stop is a customer or a vendor. Given a set of customers $A$ and vendors $V$, the set of all possible routes is given by $\mathcal{R}(V, A) = \{r_1, \ldots, r_{|\mathcal{R}|}\}$. We use *parentheses notation* in a route, $(a_i, \ldots, a_j)$ and $(v_i, \ldots, v_j)$, to denote random permutations of customers or vendors, respectively. For example, a specific route with three vendors and three customers could be $r = [(v_1, v_2, v_3), a_3, (a_1, a_2)]$, and the parentheses indicate that the drone may pick up orders from vendors $v_1$, $v_2$, and $v_3$ in any order, then drop off a package at customer $a_3$, and finally drop off packages at customers $a_1$ and $a_2$ in any order.

The term *runs* refers to consecutive vendor or customer visits, and $n_v \leq |V|$ and $n_a \leq n$ refer to the lengths of a vendor run and a customer run, respectively. Visits to vendors and customers in sequential order may be denoted correspondingly with $v_{i:j}$ (meaning $v_i, v_{i+1}, \ldots, v_{j-1}, v_j$) and $a_{i:j}$ (meaning $a_i, a_{i+1}, \ldots, a_{j-1}, a_j$). For example, the route given above contains a vendor run of $n_v = 3$ and a customer run of $n_a = 3$ and could also be written as $[(v_{1:3}), a_3, (a_{1:2})]$.

## III. PRIVACY RISK: DEFINITION AND COMPUTATION

*A. Defining privacy risk*

To define the privacy risk for an individual customer $a_i$, we first note that given a set of customers, vendors, and fixed route, a third-party observer can compute the probabilities that a customer $a_i$ received a package from each vendor. We show this with an example.

Consider the route $[v_1, v_2, a_2, v_3, a_3, a_1]$ where $v_1$, $v_2$, and $v_3$ are vendors supplying apples, balls, and pens, respectively, and customers $a_1$, $a_2$, and $a_3$ ordered an apple, a ball, and a pen, respectively. This route is illustrated on the left side of Fig. 2. A third-party observer sees the vendors and customers, but does not know which vendor delivered to which customer (right side of Fig. 2). We provide a stop-by-stop analysis from the third-party observer's perspective in Table I.

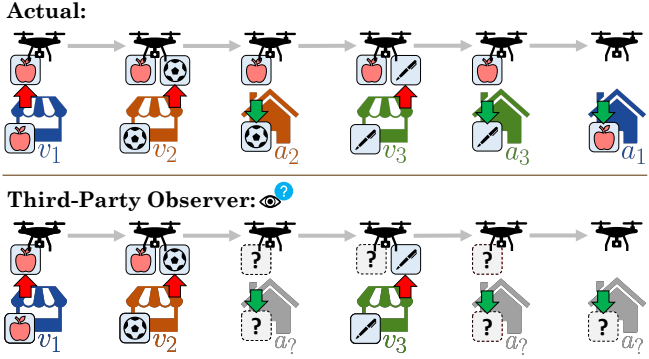

Fig. 2. Above: The true route and orders. Below: The route seen by the third-party observer, which must reason about product assignments. Up and down arrows represent pick-ups and drop-offs, respectively.

For $a_i \in \{a_1, a_2, a_3\}$ and $v_j \in \{v_1\text{:apple}, v_2\text{:ball}, v_3\text{:pen}\}$, let $p_{a_i}(v_j)$ be the probability that the item from $v_j$ is dropped off at $a_i$. Recalling that $a_1$, $a_2$, and $a_3$ ordered the apple, ball, and pen, respectively, we can compute $p_{a_i}$ by marginalizing the last row of Table I over all $a_j \neq a_i$. The resulting probability distributions for $a_1$, $a_2$, and $a_3$ are shown in Fig. 3. More generally, if we have a set of vendors $V = \{v_1, \ldots, v_m\}$, we define $p_{a_i} : V \to [0, 1]$ as the probability mass function that assigns the probability of $a_i$ receiving an item from $v_j$ for each vendor $v_j \in V$.

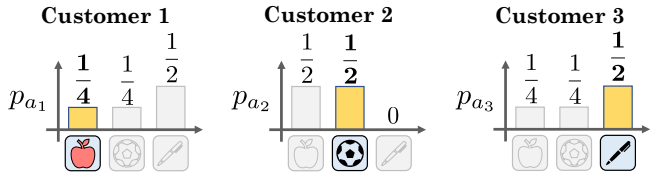

Fig. 3. Probability mass functions of vendors for customers, with probability of determining correct item assignment highlighted.

TABLE I
STOP-BY-STOP ANALYSIS OF CUSTOMER ORDER PROBABILITIES FROM A THIRD-PARTY OBSERVER'S PERSPECTIVE, TRAVERSING THE ROUTE TOP-DOWN FROM THE FIRST ROW.

| Stop | Action | Item | Customer Orders | | | Payload | Prob. |
|---|---|---|---|---|---|---|---|
| | | | $a_1$ | $a_2$ | $a_3$ | | |
| $v_1$ | Pick-up | Apple | ? | ? | ? | {Apple} | 1 |
| $v_2$ | Pick-up | Ball | ? | ? | ? | {Apple, Ball} | 1 |
| $a_2$ | Drop-off | ? | ? | Apple | ? | {Ball} | 1/2 |
| | | | ? | **Ball** | ? | {Apple} | 1/2 |
| $v_3$ | Pick-up | Pen | ? | Apple | ? | {Ball, Pen} | 1/2 |
| | | | ? | **Ball** | ? | {Apple, Pen} | 1/2 |
| $a_3$ | Drop-off | ? | ? | Apple | Ball | {Pen} | 1/4 |
| | | | ? | Apple | **Pen** | {Ball} | 1/4 |
| | | | ? | **Ball** | Apple | {Pen} | 1/4 |
| | | | ? | **Ball** | **Pen** | {Apple} | 1/4 |
| $a_1$ | Drop-off | ? | Pen | Apple | Ball | empty | 1/4 |
| | | | Ball | Apple | **Pen** | empty | 1/4 |
| | | | Pen | **Ball** | Apple | empty | 1/4 |
| | | | **Apple** | **Ball** | **Pen** | empty | 1/4 |

Note that in the above example, the choice of route $r$ directly impacts $p_{a_i}$. Thus, given a fixed route $r \in \mathcal{R}(V, A)$ and a selected customer $a_i \in A$, we denote this route-conditioned probability mass function by $p_{a_i}(v_j \mid r)$, which returns a valid probability for each $v_j \in V$ given a customer $a_i$ and route $r$. Explicitly, $p_{a_i}(v_j \mid r)$ is the probability to a third-party observer that vendor $v_j$ delivered a package to customer $a_i$ in route $r$.

Then, our definition of privacy risk follows directly:

**Definition 1** (Privacy risk). *Given a route $r \in \mathcal{R}(V, A)$, the privacy risk $\Delta(i, r)$ to customer $i$ is given by*

$$\Delta(i, r) = p_{a_i}(v_i \mid r). \tag{1}$$

For brevity, we write $\Delta(i)$ when the route $r$ is fixed or clear from context. Additionally, we denote a vector of privacy risks by $\boldsymbol{\Delta} = (\Delta(1), \ldots, \Delta(n))^\intercal$, with the understanding that it is specific to a route $r$. Given a route, Algorithm 1 allows us to explicitly compute $\boldsymbol{\Delta}$ without constructing the entire probability mass function for each $a_i$.

We now provide two route-level privacy risk measures that we use throughout the remainder of this paper:

**Definition 2** (Worst-case privacy risk). *Given a route $r \in \mathcal{R}(V, A)$, the worst-case privacy risk $\Delta_{\max}(r)$ is given by*

$$\Delta_{\max}(r) = \max_{i \in \{1, \ldots, n\}} \Delta(i, r). \tag{2}$$

**Definition 3** (Average privacy risk). *Given a route $r \in \mathcal{R}(V, A)$, the average privacy risk $\overline{\Delta}(r)$ of $r$ is given by*

$$\overline{\Delta}(r) = \frac{1}{n} \sum_{i=1}^{n} \Delta(i, r). \tag{3}$$

Again for brevity, we use $\Delta_{\max}$ and $\overline{\Delta}$ for worst-case and average privacy risk, respectively, when the route is fixed or clear from context.

### B. Computing privacy risk

We propose an algorithm to evaluate the privacy risk for each order given a route. The key insight is that lack of knowledge of what occurs at each stop leads to obfuscation of the vendor from which a customer receives an item. Moreover, because we are only concerned with the probability of *correctly* matching vendor to a customer, privacy risk reduction for an order cannot occur before the associated vendor is visited. We now provide a procedure for computing these privacy risks, which can be summarized as finding the probability that item $i$ is not dropped off prior to $a_i$ and is dropped off at $a_i$.

Consider an arbitrary part of a route consisting of a vendor run followed by a customer run. The drone begins with a $Payload$ set containing $n_{\text{pl},0} = |Payload|$ items previously picked up and not yet delivered, picks up items from a set of $n_v$ vendors, $VendorRun$, delivers items to a set of $n_a$ customers, $CustomerRun$, and then continues with additional vendor and customer stops.

After the $n_v$ vendor visits, there are $n_{\text{pl}} = |Payload| = n_{\text{pl},0} + n_v$ orders aboard the drone. For the $n_a$ customers that follow, there is a $\frac{1}{n_{\text{pl}}}$ probability of selecting the correct item *if* it is aboard the drone. To find the probability that an item remains on the drone, we can compare the payload size at the end of the customer run to that at the start of it; then, the probability that an item for a future customer was not dropped off is $\frac{|Payload|}{n_{\text{pl}}} = \frac{n_{\text{pl},0} + n_v - n_a}{n_{\text{pl},0} + n_v}$.

For example, if a drone carries a package from vendor $v_1$ ($n_{\text{pl},0} = 1$), visits $v_2, v_3, a_2$ ($n_v = 2$, $n_a = 1$), and then moves on to more vendors and customers, the privacy risk updates would be computed as follows:

- The drone starts its customer run with three packages ($n_{\text{pl}} = n_{\text{pl},0} + n_v = 3$) and drops one off at $a_2$. The privacy risk to $a_2$ is multiplied by $\frac{1}{n_{\text{pl}}} = \frac{1}{3}$. Because package 2 was just picked up, the privacy risk for order 2 is likewise $\Delta(2) = \frac{1}{3}$.
- The drone finishes its customer run with two packages (1 and 3, $n_{\text{pl},0} + n_v - n_a = 2$). Having started the run with three packages, the privacy risks of the remaining orders are multiplied by $\frac{2}{3}$, as that is the probability that the respective packages were *not* left at $a_2$. That is, $\Delta(3) = \frac{2}{3}$ and $\Delta(1) = \frac{2}{3}\Delta_0(1)$ where $\Delta_0(1)$ is the privacy risk of order 1 at the start of the vendor run.

This procedure is reflected in Algorithm 1.

IV. PRIVACY ANALYSIS

We now present some results regarding privacy risk. Note that when we reference average and worst-case privacy risk, we refer to the risk of the best possible route given the constraints, not the risk over all possible routes. Due to space limitations, we omit proposition proofs.

---

**Algorithm 1** Privacy risk computation

**Input:** Route, number of orders $n$
**Output:** Privacy risk for each order

$\Delta \leftarrow \mathbf{1}_{n \times 1}$
$Payload \leftarrow \{\}$
**for** ($VendorRun$ followed by $CustomerRun$) $\in Route$ **do**
　**for** Vendor $v_i \in VendorRun$ **do**
　　| Add package $i$ to $Payload$ (pick-up)
　**end**
　$n_{\text{pl}} \leftarrow |Payload|$
　**for** Customer $a_i \in CustomerRun$ **do**
　　$\Delta(i) \leftarrow \Delta(i)/n_{\text{pl}}$
　　Remove package $i$ from $Payload$ (drop-off)
　**end**
　$ProbMadeThrough \leftarrow |Payload|/n_{\text{pl}}$
　$\Delta(i) \leftarrow \Delta(i) \times ProbMadeThrough$ for $i \in Payload$
**end**
**return** $\Delta$

---

*A. Aggregation*

We begin our analysis with a consideration of the simplest of cases: $n$ orders where all $n$ vendors are visited first followed by all $n$ customer addresses.

**Proposition 1** (Privacy from aggregation). *The route* $r = [(v_{1:n}), (a_{1:n})]$ *has* $\Delta(i) = \overline{\Delta} = \Delta_{\max} = 1/n$.

With this route, the ability of a third-party observer to determine the vendor from which a customer received an item is no better than a random guess, or $\frac{1}{n}$. Clearly, the aggregation of orders benefits privacy, with diminishing returns as the number of orders increases.

*B. Capacity constraints*

Drones, of course, have a finite capacity, which means we cannot always obtain our desired $\frac{1}{n}$ privacy risk for all orders. Consider the case of a drone with capacity 1. The only possible route for this drone to take to fulfill an order of size $n$ is $[v_1, a_1, \ldots, v_n, a_n]$, which has maximal privacy risk $\Delta(i) = \Delta_{\max} = \overline{\Delta} = 1$, i.e., the assignment of $a_i$ to $v_i$ is obvious.

**Proposition 2** (Privacy limitation of capacity). *Any route for a drone of capacity $c$ serving $n$ orders has* $\Delta_{\max} \geq \max\left(\frac{1}{n}, \frac{1}{c}\right)$.

A capacity-constrained drone with capacity $c$ is limited to visiting at most $c$ vendors and picking up $c$ items before it must go to at least one customer. Then, the first customer visited (not necessarily $a_1$) has a privacy risk of $\frac{1}{c}$ regardless of how many total orders there are.

Like the worst-case privacy risk, the average privacy risk in an order-limited route can (only) be improved by increasing the number of orders. Unlike the worst-case privacy risk, however, the average privacy risk in a capacity-limited route may be reduced by increasing the drone's capacity *or* by increasing the number of orders the drone fulfills, as long as $c > 1$.

These results are reflected in Fig. 4. Except along the $c = 1$ and $\overline{\Delta} = \frac{1}{n}$ curves, there are two ways to reduce the average privacy risk: (i) replace the drone with one of a larger capacity (i.e., moving down to another capacity curve) or (ii) route the drone to handle more orders (i.e., moving to the right along a capacity curve). The privacy benefit of (i) is from increased aggregation, while the benefit of (ii) is because ambiguity can be increased with a greater variety of items being carried around, even if the maximum number of items carried at once remains the same.

Consider once again the route $[v_1, v_2, a_2, v_3, a_3, a_1]$ as a baseline for fulfilling three orders with a drone of capacity $c = 2$, for which $\Delta(1) = 1/4$, $\Delta(2) = \Delta(3) = 1/2$, and $\overline{\Delta} = 5/12$. Then, increasing the capacity of the drone from $c = 2$ to $c = 3$ enables the route $[v_1, v_2, v_3, a_1, a_2, a_3]$, which has $\Delta(1) = \Delta(2) = \Delta(3) = \overline{\Delta} = 1/3$, a $1/12$ improvement over the baseline; this is an example of (i). Alternatively, adding another order to the route for a drone of capacity $c = 2$ can also improve privacy, such as with the route $[v_1, v_2, a_2, v_3, a_1, v_4, a_3, a_4]$, which has $\Delta(1) = \Delta(3) = 1/4$, $\Delta(2) = \Delta(4) = 1/2$, and $\overline{\Delta} = 3/8$, a $1/24$ improvement over the baseline; this is an example of (ii).

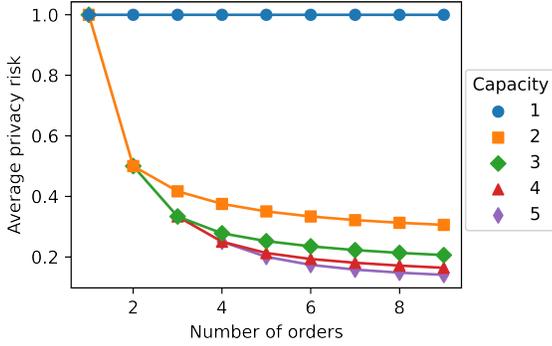

Fig. 4.    Effects of order count and drone capacity on average privacy risk.

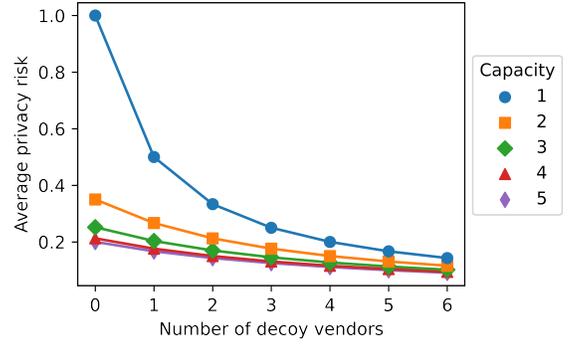

Fig. 5.    Effects of decoy stops and capacity on average privacy risk for $n = 5$ orders.

*C. Decoy vendors*

Increasing drone capacity and adding orders are not the only ways to improve privacy. As suggested previously in Example 4, we can include decoy vendors to help obfuscate vendor-customer connections. Note that decoy stops at residential addresses that did *not* place delivery orders can also improve privacy, but we only consider decoy stops at vendors. This is because individuals at potential decoy residential addresses may not appreciate drones randomly and unexpectedly loitering or landing, and decoy residential addresses lack the potential utilities of stops at decoy vendors (e.g., drone charging).

**Proposition 3** (Privacy improvement from decoy vendors). *Any route for a drone of capacity $c$ serving $n$ orders and visiting up to $n_d$ decoy vendors has $\Delta_{\max} \geq \max\left(\frac{1}{n+n_d}, \frac{1}{c+n_d}\right)$.*

Decoy vendors reduce average privacy risk: For a drone of any capacity, average privacy risk may be reduced arbitrarily with enough decoy vendors, with diminishing returns for each one. Furthermore, for a drone of capacity $c < n$, decoy vendors can aid in closing the gap in average privacy risk compared to a drone of capacity $c' \geq n$. That is, as the number of decoy vendors increases, the privacy advantage of a larger drone over a smaller one tends to zero. We show these effects in Fig. 5 for a fixed $n = 5$.

*D. Privacy and efficiency trade-offs*

Having presented several techniques to improve privacy—aggregating orders, increasing drone capacity, and visiting decoy vendors—we now consider the potential cost of these privacy benefits. We choose to analyze the cost from an efficiency perspective as measured by the average amount of time customers are required to wait for their orders.

**Definition 4** (Wait time and average wait time). *Given a route $r \in \mathcal{R}(V, A)$ and customer $a_i$, define the* wait time $W(i, r)$ *to be the amount of time $a_i$ must wait before receiving its order. Accordingly, the* average wait time $\overline{W}(r)$ *of $r$ is given by*

$$\overline{W}(r) = \frac{1}{n} \sum_{i=1}^{n} W(i, r). \tag{4}$$

For brevity, we write $W(i)$ and $\overline{W}$ for wait time and average wait time, respectively, when the route $r$ is fixed or clear from context.

We find that the trade-off between privacy and efficiency is heavily impacted by geography; in some situations, a route may be optimal in both privacy *and* efficiency, whereas other settings require a choice between either privacy *or* efficiency. By way of example, consider the cases in Fig. 6, where two vendors and two customers are at the corners of a unit square. For simplicity, we assume the drone begins at $v_1$ and use the distance traveled prior to visiting $a_i$ as a proxy for wait time. We also ignore the time taken to pick up or drop off items.

The privacy and efficiency analyses for these routes are shown in Table II. In the situation with routes $r_{ul}$ and $r_{ll}$, privacy and efficiency can be simultaneously improved with a drone of sufficient capacity. On the other hand, for $r_{ur}$ and $r_{lr}$, only one of privacy or efficiency may be increased at the expense of the other.

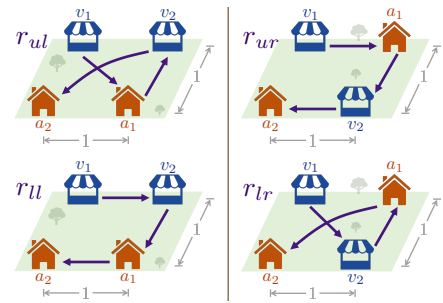

Fig. 6.    Left: A setting in which privacy and efficiency can be simultaneously improved. Right: A setting in which privacy comes at the cost of efficiency.

TABLE II
PRIVACY AND EFFICIENCY ANALYSIS FOR THE FOUR ROUTES IN FIG. 6
WHERE $r_{ul}, r_{ll}, r_{ur}$, AND $r_{lr}$ REFER TO THE UPPER LEFT, LOWER LEFT,
UPPER RIGHT, AND LOWER RIGHT ROUTES, RESPECTIVELY.

| $r$ | $\Delta(1)$ | $\Delta(2)$ | $\Delta_{\max}$ | $\overline{\Delta}$ | $W(1)$ | $W(2)$ | $\overline{W}$ |
|---|---|---|---|---|---|---|---|
| $r_{ul}$ | 1 | 1 | 1 | 1 | 1.414 | 3.828 | 2.621 |
| $r_{ll}$ | 1/2 | 1/2 | **1/2** | **1/2** | 2 | 3 | **2.5** |
| $r_{ur}$ | 1 | 1 | 1 | 1 | 1 | 3 | **2** |
| $r_{lr}$ | 1/2 | 1/2 | **1/2** | **1/2** | 2.414 | 3.828 | 3.121 |

## V. REAL-WORLD CASE STUDIES

We now consider how real-world geography influences the trade-off between privacy and efficiency. However, to do so at scale entails solving the traveling salesman problem and thus is computationally intractable. Moreover, as discussed in Section IV-D, geography and efficiency complicate the valuation of privacy. Therefore, we are interested in routing heuristics that can give good, albeit sub-optimal, routes near the Pareto front of the privacy-efficiency trade-off.

Our heuristic routes do require permuting the order of customer visits within parentheses to determine the most efficient route. However, they prevent the need to enumerate all possible routes *and* the ordering of customers within those routes. In the forthcoming case studies, we evaluate our heuristics against the Pareto front found by rote enumeration of all possible routes for $n = 6$ orders.

### A. Routing heuristics

We propose some heuristics for generating routes with specific privacy properties in Table III; due to limitations of space, we omit the proofs of their privacy risks. For a route with $n$ orders, we denote the route given in Proposition 1, i.e., $n$ vendor visits followed by $n$ customer visits, an $n$-tuple. A $k$-$l$ split is a $k$-tuple followed by an $l$-tuple such that $k + l = n$, while an $n$-$k$ reversal, where $n \geq 2k$, is an $n$-tuple with the last $k$ vendors and the first $k$ customers exchanged. Finally, $c$-stuffing is a heuristic for a drone of capacity $c$, with average privacy risk tending to $\overline{\Delta} = \frac{1}{c}\left(\frac{c-1}{c}\right)^{c-1}$ as $n$ grows large.

From our experiments with real-world geographies, we find heuristic properties to be consistent with our analysis. All $k$-$l$ splits have the same average privacy risk, which aligns with the fact that $\overline{\Delta} = \frac{2}{k+l} = \frac{2}{n}$ for $k$-$l$ splits. Larger drones may be more efficient than smaller drones because they have the option—but not the obligation—to partition orders differently from drones of smaller capacity. Results from the $k$-reversal heuristic match Section IV-B, with drones of larger capacity having lower average privacy risk. However, the lower average privacy risk from a larger drone comes at the cost of efficiency in most geographies. We also observe that $c$-stuffing is optimal in privacy for a fixed capacity but may not yield the most efficient route; we posit this is because the heuristic takes a "first-in, first-out" approach, so no order is delivered too soon after it is picked up, which would reduce uncertainty and increase privacy risk.

### B. Impacts of geographical distribution

We analyze geographical influences on the trade-off between privacy and efficiency using four realistic delivery scenarios with $n = 6$, all shown in Fig. 7. We use real distances from the Boston metropolitan area, assume that a drone has an average speed of 20 m/s, and let stops take one minute each. We also assume vendors are visited in order from $v_1$ to $v_6$ with no decoy vendors. To be clear, we are more concerned with the qualitative impacts of geography than with finding the most efficient or the most private route.

*1) Uniform:* The upper left of Fig. 7 depicts a map in which vendors and customers are uniformly distributed. In this situation, there is a limit to the efficiency improvements that increasing drone capacity can provide. Above $c = 3$, additional capacity can only improve privacy at the expense of efficiency. To understand why this may be, consider a drone of capacity $c = 6$ that aims to maximize privacy. Doing so would require visiting all vendors before visiting any customers. Then, even if there is a customer between two vendors whose package has already been picked up, the drone cannot stop to drop off that package and must instead backtrack after visiting all vendors.

*2) Two clusters:* The two clusters map in the upper right of Fig. 7 has all vendors in close proximity to one another and well-separated from similarly-clustered customers. The benefits of capacity are most apparent in this topology: A larger drone can improve privacy via aggregation and also improve efficiency by taking fewer trips between the two clusters. Additionally, we find that with a limited capacity, the split heuristic yields more efficient routes while the reversal heuristic yields more private routes.

*3) Hub and spoke:* The hub and spoke map in the lower left of Fig. 7 may be representative of an area in which vendors are clustered in the city center and are surrounded by outlying residential areas. Qualitatively, the benefits of increasing drone capacity are more similar to those of the map with two clusters than the uniform map, but the split and reversal heuristics are not as optimal as in the two cluster case.

*4) Linear:* The lower right of Fig. 7 shows a linear map, which could be representative of an urban area containing a main thoroughfare street with parallel vendors and customers. Here, the reversal and stuffing heuristics struggle with efficiency because they are forced to visit later vendors before backtracking to earlier addresses.

TABLE III
ROUTING HEURISTICS AND ASSOCIATED REQUIRED CAPACITY, PRIVACY RISK $\Delta(i)$, WORST-CASE PRIVACY RISK $\Delta_{\max}$, AND AVERAGE PRIVACY RISK $\overline{\Delta}$.

| Heuristic | Route | Capacity | $\Delta(i)$ | $\Delta_{\max}$ | $\overline{\Delta}$ |
|---|---|---|---|---|---|
| $k$-$l$ split | $(v_{1:k}),(a_{1:k}),\ldots$ $\ldots,(v_{k+1:k+l}),(a_{k+1:k+l})$ | $\max(k,l)$ | $\begin{cases}\frac{1}{k} & \text{if } 1\leq i\leq k \\ \frac{1}{l} & \text{if } k+1\leq i\leq k+l\end{cases}$ | $\max\left(\frac{1}{k},\frac{1}{l}\right)$ | $\frac{2}{k+l}$ |
| $k$-reversal | $(v_{1:n-k}),(a_{1:k}),\ldots$ $\ldots,(v_{n-k+1:n}),(a_{k+1:n})$ | $n-k$ | $\begin{cases}\frac{1}{n-k} & \text{if } 1\leq i\leq k \text{ or } n-k+1\leq i\leq n \\ \frac{n-2k}{(n-k)^2} & \text{if } k+1\leq i\leq n-k\end{cases}$ | $\frac{1}{n-k}$ | $\frac{n^2-2nk+2k^2}{n(n-k)^2}$ |
| $c$-stuffing | $(v_{1:c}),a_1,v_{c+1},a_2,\ldots$ $\ldots,v_n,(a_{n-c+1:n})$ | $c$ | $\frac{1}{c}\left(\frac{c-1}{c}\right)^{\min(c-1,n-c,i-1,n-i)}$ | $\frac{1}{c}$ | $\dfrac{2\left[\sum_{j=0}^{d}b^j\right]+(n-2(d+1))b^d}{nc}$ where $b=(c-1)/c$ and $d=\min(c-1,n-c)$ |

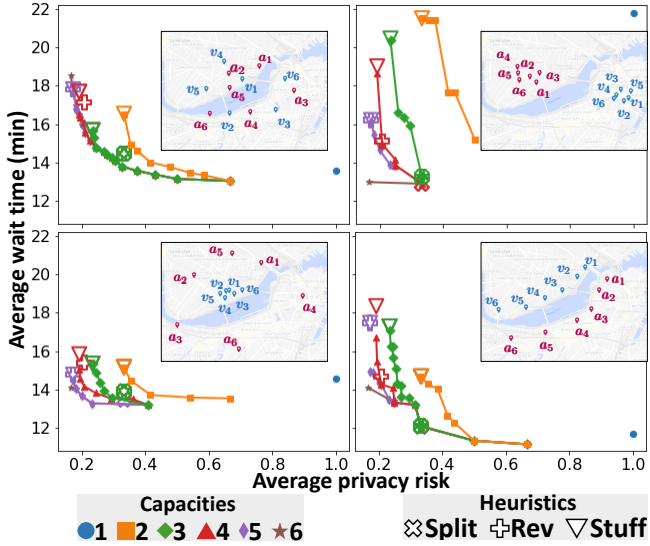

Fig. 7. Privacy and efficiency trade-offs and heuristic performance with drones of various capacities. Upper left: uniform map. Upper right: map with two clusters. Lower left: hub and spoke map. Lower right: linear map. Note that some heuristics overlap with each other in the trade-off plots.

## VI. CONCLUDING REMARKS

In this work, we addressed an emerging business model and consequential privacy risks of drone package delivery services. We provided a framework and algorithm to compute these privacy risks and suggested strategies to mitigate them. However, in doing so, we made many simplifying assumptions. For example, we assumed drone capacity constraints were uniform across orders, whereas in reality different products occupy varying amounts of cargo space. Also, a key assumption for our privacy risk calculations was that each order was from a different vendor. Furthermore, our analysis considered privacy for a customer based on the routing for just one of their orders, even though third-party observers may collect, aggregate, and analyze route and trajectory data from multiple deliveries across time; future work in this direction should define and address such associated longitudinal privacy risks from orders over time. Future work can also address our other limiting assumptions by extending the privacy risk analysis and routing strategies to consider multiple orders from the same vendor as well as grouping vendors by product or service type, the kind of information that may be inferred about their customers, or the level of risk that may be associated with privacy losses. Finally, the routing heuristics we provided may be extended to online routing problems, where additional orders are added dynamically to in-progress delivery routes.

## REFERENCES

[1] C. Mehrtens, "Novant Health launches drone operation for COVID-19 response," 2020, accessed: October 2021.
[2] R. de León, "Role of medical drones in global Covid vaccine campaign is growing," 2021, accessed: October 2021.
[3] T. Ward, "Walmart and Zipline Team Up to Bring First-of-Its Kind Drone Delivery Service to the United States," 2020, accessed: October 2021.
[4] B. Kruk, "The sky's no longer the limit," 2019, accessed: January 2022.
[5] M. McFarland, "Google drones will deliver Chipotle burritos at Virginia Tech," 2016, accessed: October 2021.
[6] "Wing," https://wing.com, accessed: February 2022.
[7] "Manna drone delivery," https://www.manna.aero, accessed: February 2022.
[8] Federal Aviation Administration, "86 FR 4390: Remote Identification of Unmanned Aircraft," 2021, accessed: October 2021.
[9] ——, "UAS Remote Identification Overview," 2021, accessed: January 2022.
[10] A. Berke, N. Lee, and P. Chwalek, "Private delivery networks," *arXiv preprint arXiv:2108.07354*, 2021.
[11] F. Brunton and H. Nissenbaum, "Vernacular resistance to data collection and analysis: A political theory of obfuscation," *First Monday*, 2011.
[12] H. Nissenbaum and H. Daniel, "Trackmenot: Resisting surveillance in web search," 2009.
[13] "The topics api - chrome developers," https://developer.chrome.com/docs/privacy-sandbox/topics, accessed: February 2022.
[14] J. Karlin, "The topics API," https://github.com/jkarlin/topics, accessed: February 2022.